\documentclass{article}
\usepackage{spconf,amsmath,graphicx}
\usepackage{xcolor}
\usepackage{subcaption}
\usepackage{booktabs}
\usepackage{array}
\usepackage{xcolor}
\usepackage{hyperref}

\newcolumntype{x}[1]{>{\centering\arraybackslash\hspace{0pt}}p{#1}}

\title{STREAMING TARGET SOUND EXTRACTION USING ATTENTION QUERIES}

\title{Causal TARGET SOUND Selection USING ATTENTION QUERIES}

\title{Transformers for causal label-based source extraction}

\title{Real-time TARGET SOUND EXTRACTION USING ATTENTION QUERIES}

\title{Causal Target source extraction using a transformer decoder}

\title{Real-time label-based  sound extraction}

\title{Real-time  TARGET SOUND EXTRACTION USING Transformers}

\title{Streaming TARGET SOUND EXTRACTION using attention}

\title{Real-time TARGET SOUND EXTRACTION using attention}

\title{Real-time TARGET SOUND EXTRACTION}

\title{Waveformer:  Transformer  for Real-time TARGET SOUND EXTRACTION}

\title{Real-time TARGET SOUND EXTRACTION}

\title{Real-time TARGET SOUND EXTRACTION}

\name{Bandhav Veluri,$^\diamond$ Justin Chan,$^\diamond$ Malek Itani,$^\diamond$ Tuochao Chen,$^\diamond$ Takuya Yoshioka,$^\bullet$  Shyamnath Gollakota$^\diamond$ }
\address{$^\diamond$Paul G. Allen School of Computer Science \& Engineering, University of Washington, USA\\ $^\bullet$Microsoft, One Microsoft Way, Redmond, WA, USA}
\newcommand{\xref}[1]{\S\ref{#1}}
\begin{document}
\ninept
\maketitle
\begin{abstract}

We present  the first neural network model to achieve  real-time and streaming target sound
extraction. To accomplish  this, {we propose \textit{Waveformer}}, an  encoder-decoder architecture with a stack of dilated causal convolution layers as the encoder, and a transformer decoder layer as the decoder. This hybrid architecture uses  dilated causal convolutions for processing large receptive fields in a computationally efficient manner, {while also leveraging the generalization performance of transformer-based architectures}. Our evaluations show as much as 2.2--3.3~dB improvement in SI-SNRi compared to the prior models for this task while having a 1.2--4x smaller model size and a  1.5--2x {lower} runtime. We provide code, dataset, and audio samples: \textcolor{blue}{\url{https://waveformer.cs.washington.edu/}}.



\end{abstract}
\begin{keywords}
Sound selection, streaming, attention 
\end{keywords}

\section{Introduction}
\label{sec:intro}
Humans are exceptionally adept at attending their auditory focus to specific sounds even in a noisy environment~\cite{2020arXiv200605712O}. Recent works that aim to create a computational equivalent of this human capability  formulate this problem as target sound extraction~\cite{2020arXiv200605712O,delcroix2021few,delcroix2022soundbeam}.  The goal   is to extract sound signals of interest from a mixture of various overlapping sounds, given clues that provide information about the target sound class such as embeddings of a one-hot label~\cite{2020arXiv200605712O}, audio clips~\cite{delcroix2022soundbeam,gfeller2021one}, and images~\cite{gao2019co,xu2019recursive}. Streaming target sound extraction  could enable real-time  intelligent acoustic  applications for  headphones, hearing aids, and telephony  by filtering out undesired sounds from the environment (e.g., traffic) and presenting only sounds of interest to the user (e.g., sirens). 


Recent works on target sound extraction  have shown promising performance even for mixtures containing a large  number of  sound classes~\cite{2020arXiv200605712O}. However, none of these  prior works   demonstrate real-time  streaming capabilities. In particular, the prior works for this task  are based on non-streaming models and designed for offline processing, where the neural network has access to a large block 
($\ge 1$~s) of audio samples~\cite{2020arXiv200605712O}. In contrast, real-time  streaming applications impose significant {algorithmic and} computational  constraints, requiring networks to     operate on small blocks 
($\le 10$~ms) with a limited number of lookahead samples  for each block. All these factors can significantly degrade  the performance~\cite{luo2019conv}.

In this paper, we present the first deep learning method to perform  target sound extraction in a streaming manner.  Fig.~\ref{fig:main_fig} shows  Waveformer, our encoder-decoder architecture where the encoder is a stack of dilated causal convolution (DCC) layers and the decoder is a transformer decoder \cite{https://doi.org/10.48550/arxiv.1706.03762}.\footnote{We call  our network, Waveformer, since it uses a hybrid architecture with the causal convolution layers,  common in WaveNet ~\cite{https://doi.org/10.48550/arxiv.1609.03499} based architectures, as the encoder and a transformer as the decoder.}  {Our intuition is that much of the complexity in  prior  models comes with processing large receptive fields, especially at high sampling rates. For example, recent transformer-based architectures proposed for speech separation \cite{subakan2022resource, luo2022tiny} implement chunk-based processing, where each chunk independently attends to all the chunks in the receptive field. Thus, to achieve a receptive field of length $R$, for each chunk, these models have an $\mathcal{O}(R)$ computational complexity. Instead, since DCC layers have a complexity of $\mathcal{O} (\log R)$ for achieving the same amount of receptive field (\xref{sec:encoder}), we use a stack of DCC layers as the encoder that processes the receptive field. We then use the decoder layer of the transformer architecture \cite{https://doi.org/10.48550/arxiv.1706.03762} as our model's decoder. {The decoder generates a mask that can extract the specified target sound to produce the output signal.}


To evaluate our network architecture, we implement a causal version of Conv-TasNet and a streaming  version of ReSepformer~\cite{subakan2022resource} for the task of streaming target sound extraction. Evaluations show that our hybrid network architecture achieves state-of-the-art performance for this task. 
Further, the smallest and largest versions of our model have real-time factors (RTFs) of  0.66 and 0.94, respectively,  on a consumer-grade {CPU}, demonstrating the real-time target sound extraction capability, thus outperforming prior models in terms of both efficiency and signal quality. 



\begin{figure*}[t!]
\centering
\begin{subfigure}[t]{.25\textwidth}
    \includegraphics[width=\linewidth]{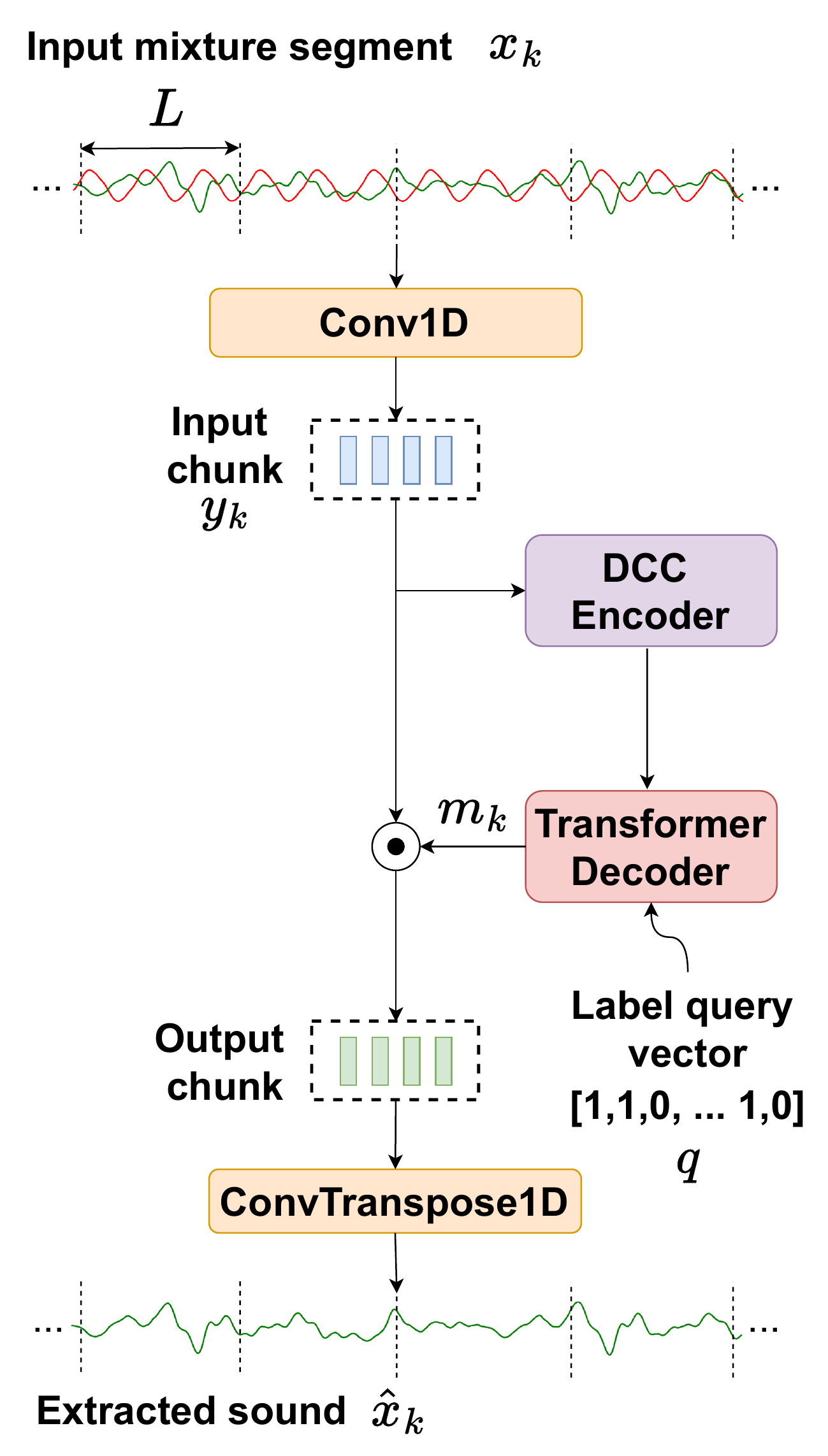}
    \label{fig:top_level}
    \caption{End-to-end architecture}
\end{subfigure}
\hfill
\centering
\begin{subfigure}[t]{.4\textwidth}
    \includegraphics[width=\linewidth]{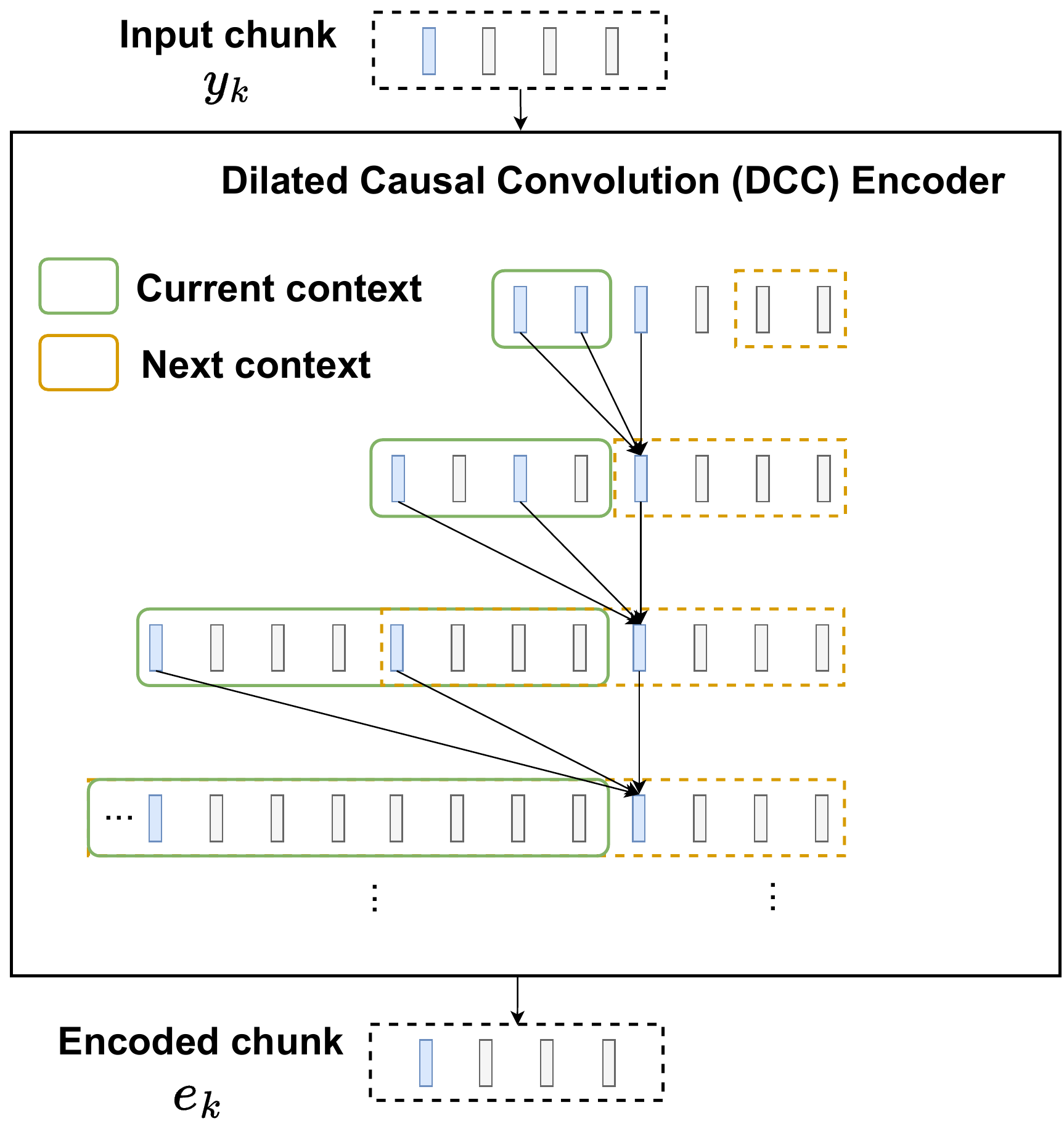}
    \label{fig:enc}
    \caption{DCC Encoder}
\end{subfigure}
\hfill
\centering
\begin{subfigure}[t]{.26\textwidth}
    \includegraphics[width=\linewidth]{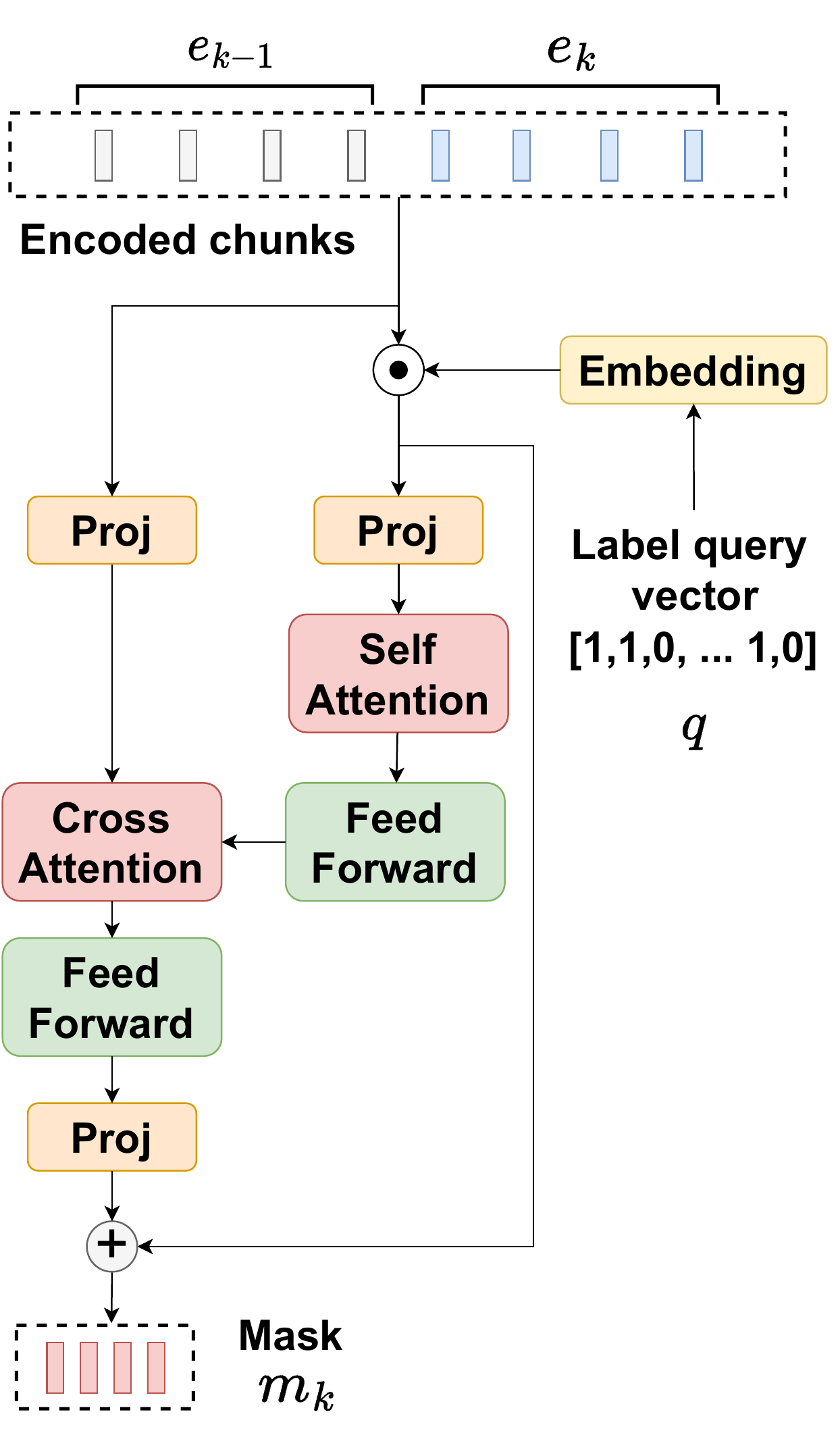}
    \label{fig:dec}
    \caption{Transformer Decoder}
\end{subfigure}
\caption{{\bf Waveformer architecture.} Streaming inference demonstrated using an example input mixture segment of length $4L$ samples, corresponding to a chunk length $K = 4$. The query is a one-hot or a multi-hot label encoding. The Dilated Causal Convolution (DCC) encoder encodes the input chunk $y_k$ using the context computed from the receptive field. The  transformer decoder computes the target mask by attending to current and previous encoded chunks.}
\vskip -0.15in
\label{fig:main_fig}
\end{figure*}

\section{Related work}
{\bf Universal sound separation.} The task here is to decompose a mixture of arbitrary sound types into their component sounds, regardless of the number of sounds in the mixture~\cite{kavalerov2019universal}. This  becomes increasingly challenging as the number of possible sound types in the mixture increases. 
Several networks have been proposed for this task including convolutional long short-term memory networks~\cite{kavalerov2019universal}, time-dilated convolution networks~\cite{kavalerov2019universal} based on Conv-TasNet~\cite{luo2019conv}, and transformer networks~\cite{zadeh2019wildmix}. Prior work also  proposed the use of embeddings learned by a sound classifier trained on a large sound ontology~\cite{gemmeke2017audio} for conditioning a separation network. 

\vskip 0.05in\noindent {\bf Target sound extraction.}
This approach can circumvent the challenge of  universal sound separation struggling to deal with a mixture of a large number of sounds. 
The clues may be provided as an embedding of an audio clip~\cite{delcroix2022soundbeam,gfeller2021one}, an image~\cite{gao2019co,xu2019recursive}, natural language text~\cite{kilgour2022text,liu2022separate}, onomatopoeic words~\cite{okamoto2022environmental}, or a one-hot sound label vector~\cite{2020arXiv200605712O}. The prior works~\cite{kong2020source,chen2022zero} have also evaluated the use of a sound event detector to detect the time when the target sound occurs in a mixture. Although these works are often motivated for practical usage~\cite{liu2022separate,2020arXiv200605712O}, none of them use streaming models. In contrast to these prior works, we design the first streaming network for target sound extraction  using attention.

\vskip 0.05in\noindent {\bf Speech-specific networks.} Prior work  has also  focused on speech enhancement~\cite{wisdom2019differentiable,eskimez2022personalized,chatterjee2022clearbuds},  speech separation~\cite{luo2019conv,luo2020dual,huang2014deep,hershey2016deep,isik2016single,yu2017permutation,wang2018voicefilter} and speech selection using clues provided to the network~\cite{tzinis2022heterogeneous,ephrat2018looking,directional,wang2022semi}. For  speech enhancement, neural networks  have been proposed~\cite{eskimez2022personalized,giri2021personalized}  to realize  real-time operation. Recently, efficient transformer-based architectures have been proposed for speech recognition and separation tasks ~\cite{subakan2021attention,luo2022tiny,https://doi.org/10.48550/arxiv.2005.08100}. These methods either use standard transformer blocks \cite{subakan2022resource} or convolution-augmented transformer blocks \cite{luo2022tiny, https://doi.org/10.48550/arxiv.2005.08100}. In contrast, we use DCC layers along with the transformer decoder. Recent work on ReSepformer~\cite{subakan2022resource} proposed a causal mode for their transformer method, which  we use  for  our baseline comparisons. 

\section{Waveformer Architecture}
 We process  individual audio chunks of duration $\tau$ seconds. For streaming, we need to operate at the chunk level: an output chunk can  depend on the current and past chunks. Thus, streaming models have an intrinsic latency equal to the duration of a single chunk.  In real-time practical systems, it is desirable that this   latency is on the order of 10 ms \cite{7952149}.  Fig.~\ref{fig:main_fig} shows our proposed time-domain model architecture, which employs an encoder-decoder  \footnote{{Our use of \emph{encoder} and \emph{decoder} is in the context of mask generation. In contrast, Conv-TasNet \cite{luo2019conv} uses those terms for input Conv1d and output ConvTranspose1D, respectively.}}based mask generation network to generate an element-wise multiplicative mask in the latent space. Let $x_k \in R^{S}$ denote the current input audio chunk, where $S = \tau F_s$ is the   number of audio samples included in the current chunk with a sampling rate of $F_s$. In the first step, a 1D-convolution layer with stride $L$ and kernel size $3L$ is applied to the input  audio chunk $x_k$ to obtain the latent space representation, $y_k \in R^{E \times K}$, where $E$ is the latent space feature dimensions and $K = \frac{S}{L}$ is the feature sequence length in the latent space. {Setting the kernel size to $3L$ and  stride  to $L$ requires an overlap of $L$ samples with the previous and the future chunk, resulting in a lookahead of {$2L$} samples. In our experiments, we set $L = 32$ samples at 44.1 kHz. This results in a lookahead of around  $1.45$~ms, which is negligible.} Given a one-hot or multi-hot query vector $q \in \{0, 1\}^{N_c}$, where $N_c$ is the total number of classes, streaming target sound extraction is achieved by computing feature masks  $m_k \in R^{E \times K}$. With the mask generation network and element-wise multiplication denoted as $\mathcal{M}$ and $\odot$, respectively, the  target sound signal, $\hat{x}_k \in R ^ S$, is computed  as:
\begin{align}
    y_k =& \text{Conv1d}(x_k), ~~~
    m_k = \mathcal{M}(y_k\ |\ y_{k-1}, .., y_2, y_1, q) \notag\\
    \hat{x}_k =& \text{ConvTranspose1d}(y_k \odot m_k).\notag
\end{align}

\subsection{Dilated causal convolution encoder}\label{sec:encoder}
Our encoder is a stack of dilated causal convolution (DCC) layers \cite{https://doi.org/10.48550/arxiv.1609.03499}, and the decoder is a transformer network \cite{https://doi.org/10.48550/arxiv.1706.03762}. The motivation for such an architecture  is that the encoder computes a \emph{contextful} representation of the input chunk, considering the previous chunks up to a certain receptive field, and the decoder conditions the encoder output with the query vector to estimate the target mask. While recent  transformer  models for speech separation \cite{subakan2021attention, subakan2022resource} have demonstrated  performance gains over convolution-based methods \cite{luo2019conv}, the latter have generally been more computationally efficient.

We attribute this efficiency gap  to the difference in the way existing transformer models  process the receptive field compared to convolution-based architectures. To achieve a receptive field of length $R$, given the chunk-based processing in existing transformer architectures, each chunk individually attends to all previous chunks in the receptive field resulting in $\mathcal{O}(R)$ complexity. In contrast, convolution based models  \cite{https://doi.org/10.48550/arxiv.1609.03499, luo2019conv} using a stack of $M$ DCC layers with kernel size $P$ and exponentially scaling dilation factors $\{2^0, 2^1, 2^2, ..., 2^{M-1}\}$ have a receptive field of $(P - 1) \cdot (2^M - 1)$. Its complexity is $\mathcal{O}(PM)$. With  a small kernel size $P$, the computational complexity of the stacked DCC layers  is $\mathcal{O}(P \cdot \log~(1 + \frac{R}{P - 1}))$ $\sim$ $\mathcal{O} (\log R)$ for it to have a receptive field of length $R$. 

 We use 10  DCC layers with a kernel size of 3 and dilation factors $\{2^0, 2^1, 2^2, ..., 2^{10-1}\}$ in our encoder, resulting in a receptive field of $(3 - 1) \cdot (2 ^ {10} - 1) = 2046 $ samples in the latent space. With the initial input convolution stride $L$ set to  0.73~ms, our encoder's receptive field  is $\approx 1.5 s$. Fig.~\ref{fig:main_fig} (b) shows an  encoding of an input chunk of length 4. For chunk-based streaming inference, the encoder maintains a context buffer for each DCC layer. This context is initially computed from the $1s$ receptive field and then updated dynamically after encoding each subsequent chunk. For encoding a chunk, each DCC layer is fed with the output chunk of the previous layer, left padded with the context of length twice the layer's dilation. After encoding a chunk, the context is updated with the rightmost elements of the padded input for it to be used in encoding the next chunk. For each input chunk $y_k$, the DCC encoder computes an encoded representation,  $e_k \in R^{E \times K}$.

\subsection{Query-conditioned transformer decoder}
\label{sec:decoder}

To get the mask, the encoded representation computed above must be conditioned with the query, $q$. To this end, we first compute an embedding, $l \in R^{E \times 1}$, corresponding to $q$. This is achieved by using an embedding layer comprising three 512-dimensional feed-forward sub-layers with an $N_C$-dimensional input and  an $E$-dimensional output. Our transformer decoder conditions the encoded chunk $e_k$ with the query embedding $l$ and derives the mask as follows. 

Fig.~\ref{fig:main_fig} (c) shows our decoder architecture. 
First, we perform multiplicative query integration~\cite{2020arXiv200605712O, delcroix2022soundbeam} to compute the conditioned representation: ${e_k}^\prime = e_k \odot l$. Since transformers are more computationally expensive with higher dimensionality, we first project the encoded representations, ${e_k}^\prime$ and $e_k$, to the decoder dimensions $D \le E$ with $1\times1$ convolution. This results in projected encoded representations,  ${pe}_k, {{pe}_k}^\prime \in R^{D \times K}$. The decoded representations are then computed by passing ${{pe}_k, {pe}_k}^\prime$ to the transformer decoder layer's self-attention and cross-attention blocks, respectively, to obtain target mask ${pm}_k \in R^{D \times K}$ in the projected decoder space. It is then projected back to the encoder dimensions with another $1\times1$ convolution layer to obtain  ${m_k}^\prime \in R^{E \times K}$. Since the bottleneck caused by the projection layers might affect the flow of gradients, 
as depicted in the diagram, we use a skip connection immediately after the multiplicative query integration to the output of the projection layer to compute the final mask: $m_k = {m_k}^\prime + {e_k}^\prime$. 

{Within the decoder, we use the    chunk-based streaming attention scheme proposed in \cite{9413535}. As shown in  Fig.~\ref{fig:main_fig}(c), for decoding the current chunk, $e_k$, the transformer decoder only  attends to the samples in the current chunk, $e_k$, and one previous chunk,  $e_{k-1}$. This ensures that the input length to the transformer decoder is fixed at $2K$ (current chunk + one previous chunk) and prevents the inference time from growing as the input audio length increases.}


\begin{table}
\caption{Performance on the single-target extraction task. In our model, $E$ and $D$ correspond to encoder and decoder dimensionalities, respectively. RTF is the real-time factor for a consumer  CPU.}
\vspace{-.8em}
\centering
\begin{tabular}{lccc}
\toprule
Model                   & Model size & RTF  & SI-SNRi  \\
\midrule
Conv-TasNet              & 4.57M      & 1.34 & 6.14    \\
ReSepformer             & 13.24M     & 1.60 & {7.26}    \\
Ours ($E = 256;~D = 128$) & 1.10M      & 0.66 & 9.02    \\
Ours ($E = 256;~D = 256$) & 1.69M      & 0.75 & 9.40    \\
Ours ($E = 512;~D = 128$) & 3.29M      & 0.88 & 9.26    \\
Ours ($E = 512;~D = 256$) & 3.88M      & 0.94 & \textbf{9.43}    \\
\bottomrule
\end{tabular}
\label{table:single_target}
\end{table}

\begin{table}[t!]
\caption{SI-SNRi  comparison in the multi-target extraction task.}
\vspace{-.8em}
\centering

\begin{tabular}{l  c c c c}
\toprule
 & \multicolumn{3}{c}{\# selected classes} \\
Model                   & 1 & 2  & 3 & Mean \\
\midrule
Conv-TasNet              & 7.20 & 3.63 & 0.19 & 3.67 \\
ReSepformer              & 7.42 & 3.56 & 0.33 & 3.77 \\
Ours ($E = 256;~D = 128$)  & 9.06 & 4.78 & 1.51 & 5.11 \\
Ours ($E = 256;~D = 256$)  & 9.12 & 4.76 & 1.31 & 5.06 \\
Ours ($E = 512;~D = 128$)  & 9.39 & 4.92 & 1.39 & 5.23 \\
Ours ($E = 512;~D = 256$)  & 9.29 & 4.92 & 1.35 & 5.19 \\
\bottomrule
\end{tabular}
\vskip -0.15in
\label{table:multi_target}
\end{table}

\section{Experiments and results}



{\bf Dataset.} We use a synthetic sound mixture dataset created from the  FSD Kaggle 2018 dataset~\cite{Fonseca2018_DCASE}. FSD Kaggle 2018 is a set of sound event and class label pairs, with 41 different sound classes, which are a subset of the Audioset ontology \cite{gemmeke2017audio}. Our synthetic dataset consists of 50k training samples, 5k validation samples, and 10k test samples. {Sound mixtures are created using the Scaper toolkit \cite{8170052} with FSD Kaggle 2018 and TAU Urban Acoustic Scenes 2019 ~\cite{Mesaros2018_DCASE} as foreground and background sources, respectively}. Foreground sound classes are randomly sampled without replacement so that each sample has 3-5 unique classes. We construct the sound mixtures by sampling 3-5s crops from each foreground sound and then pasting them on a 6s background sound. The SNRs of the foreground sounds are randomly chosen between 15 and 25 dB, relative to the background sound. Our training and validation data are sampled from the development splits of FSD Kaggle 2018 and TAU Urban Acoustic Scenes 2019, while our test samples are  from the test splits.
From each mixture, up to 3 foreground sounds are randomly selected as targets.
During training, the choices of the target foreground sounds in the training set are randomized. Since we mainly consider human listening applications for streaming target sound extraction, we run  our experiments at a 44.1 kHz  sampling rate  to cover the full audible range.

\begin{figure}[t!]
\centering
\begin{subfigure}[l]{\linewidth}
    \includegraphics[width=\linewidth]{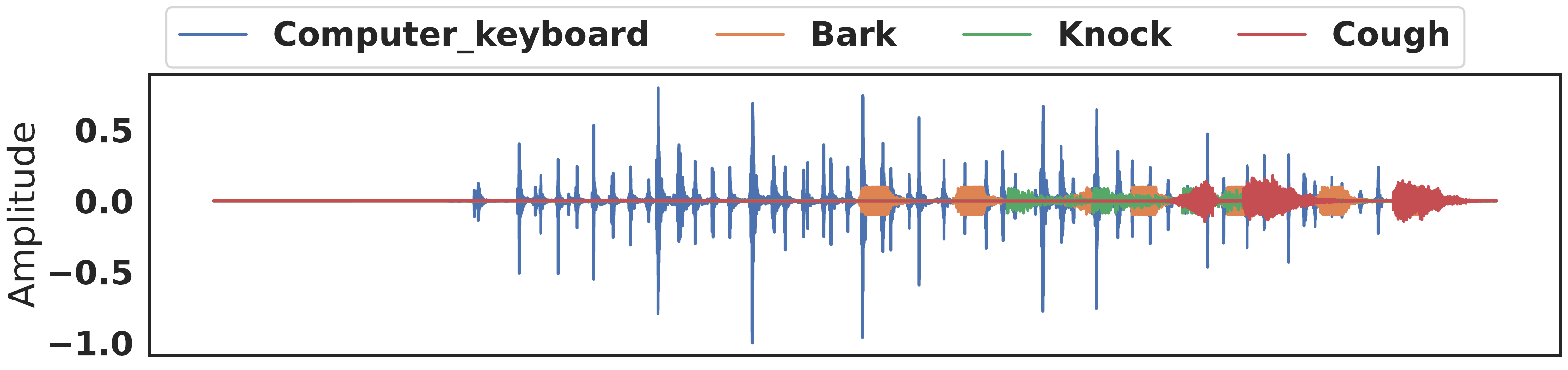}
    \caption{Input mixture}
    \label{fig:qual_mixture}
\end{subfigure}
\centering
\begin{subfigure}[l]{\linewidth}
    \includegraphics[width=\linewidth]{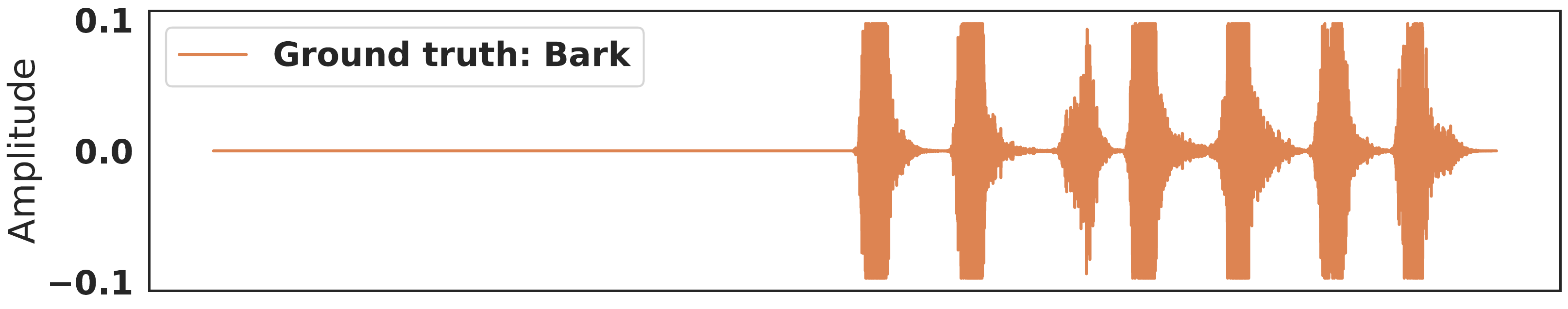}
    \caption{Ground-truth for label `Bark'}
    \label{fig:qual_gt_bark}
\end{subfigure}
\centering
\begin{subfigure}[l]{\linewidth}
    \includegraphics[width=\linewidth]{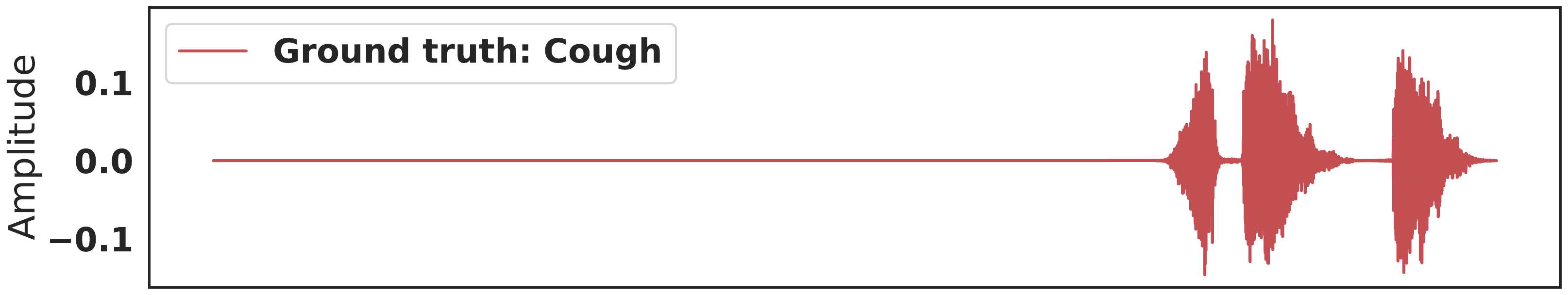}
    \caption{Ground-truth for label `Cough'}
    \label{fig:qual_gt_cough}
\end{subfigure}
\centering
\begin{subfigure}[l]{\linewidth}
    \includegraphics[width=\linewidth]{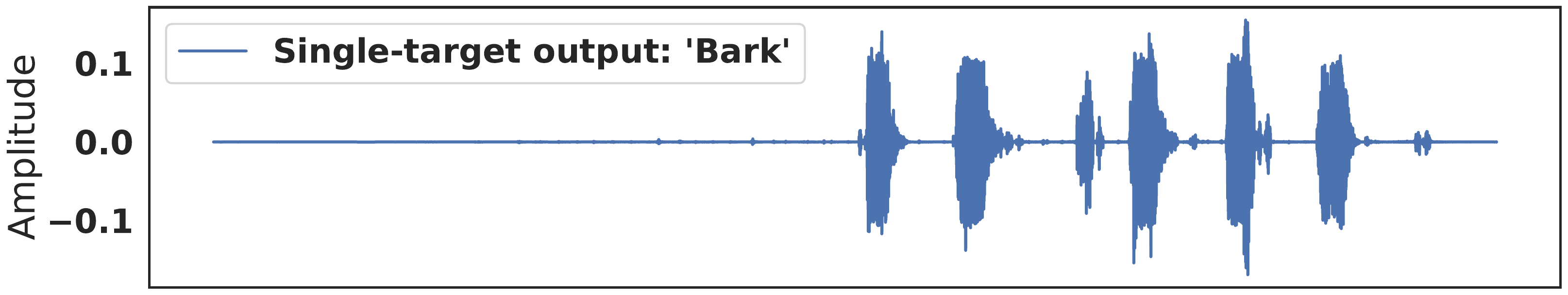}
    \caption{Output of the single-target extraction of `Bark'}
    \label{fig:qual_output_single}
\end{subfigure}
\centering
\begin{subfigure}[l]{\linewidth}
    \includegraphics[width=\linewidth]{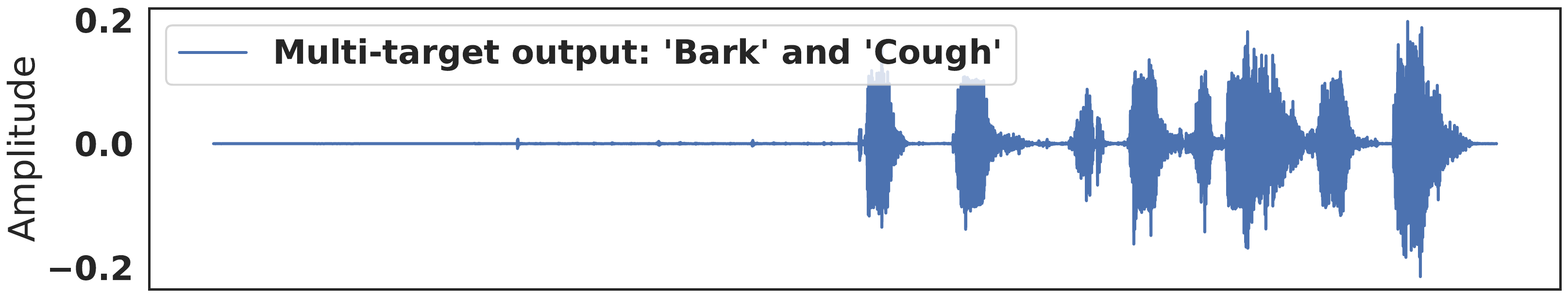}
    \caption{Output of the multi-target extraction of `Bark' and `Cough'}
    \label{fig:qual_output_multi}
\end{subfigure}
\caption{{Visualization of time-domain waveforms of single-target and multi-target extraction (x-axis represents time). }}
\label{fig:qual}
\vskip -0.2in
\end{figure}

\vskip 0.05in\noindent{\bf Evaluation setup.} Prior works \cite{2020arXiv200605712O, delcroix2022soundbeam} show that Conv-TasNet,   originally proposed for speech separation, can also be used for target sound extraction. Further,   ReSepformer proposes an efficient transformer architecture for speech separation that allows a streaming inference. Here, we compare the performance of our architecture  with the causal or streaming implementations of Conv-TasNet and ReSepformer as described in the original papers~\cite{luo2019conv,subakan2022resource} for the  target sound extraction task. 


For all the models, we set the stride of the initial convolution, $L$, to 32, which is  about $0.73$ ms at 44.1 kHz. We train multiple configurations of our model with different encoder and decoder dimensions. We fix the number of DCC layers to 10, the number of transformer layers to 1, and the chunk length, $K$, to 13. This chunk length corresponds to 416 samples in the time domain or a chunk duration of 9.43 ms.   
For Conv-TasNet, we follow the configuration used in \cite{2020arXiv200605712O} except for the number of repeats, which we set to 2. This ensures that the runtime of the Conv-TasNet baseline is not too large compared with that of our model's largest configuration. For the ReSepformer baseline, we set the model dimensionality to 512, the number of blocks to 2, the number of transformer layers to 2, and the chunk size to 13 (9.43 ms). We perform label integration after the first transformer block, as we found it to perform better than integrating it  at the beginning.

\vskip 0.05in\noindent{\bf Loss function and training hyper-parameters.} { We use a linear combination of 90\% signal-to-noise-ratio (SNR) and 10\% scale-invariant-signal-to-noise-ratio (SI-SNR) \cite{https://doi.org/10.48550/arxiv.1811.02508} as the loss function for training.
We set the initial learning rate to 5e-4 for our models and Conv-TasNet, and to 1.5e-4 for  ReSepformer. We train the models for 100 epochs and choose the model after the epoch that resulted in the best validation SI-SNRi.
}

\vskip 0.05in\noindent{\bf Results.} We separately train the models for single-target and multi-target extraction tasks and evaluate them on our testset. For multi-target evaluation, we train our model as well as baselines to make predictions with multi-hot query vectors, as opposed to one-hot queries used in the single-target evaluation. { During the multi-target training, 1-3 foreground sounds are randomly selected as target sounds. This training method using an arbitrary number of target sources helps the model learn multi-target embeddings. The same model configurations are used for both the single-target and multi-target experiments.  The Conv-TasNet and ReSepfromer baselines are also trained in the same way for the multi-target extraction task.

We also evaluate the real-time factors (RTFs) of the models on an Intel Core i5 CPU using a single thread. RTF is computed by measuring the runtime consumed by the models to process a 416 sample audio chunk (9.43 ms at 44.1 kHz), and dividing that by the chunk duration, 9.43 ms. {For the RTF measurement, we include the padding for dilated convolution layers in our model's DCC encoder and Conv-TasNet's Temporal Convolution Network (TCN) blocks, accounting for the entire receptive field. In the case of the ReSepformer, using a single chunk for RTF measurement excludes the overhead caused by causal attention masking in its inter-attention blocks. Consequently, the RTF value reported for ReSepformer is a lower bound of what is practically achievable.}}

Table~\ref{table:single_target} compares our models with different configurations with the baselines in terms of both efficiency and performance. We show that our approach  results in 2.2-3.3 dB SI-SNRi improvement compared with the baselines while being 1.5-2x more computationally efficient with 1.2-4x fewer parameters. Table~\ref{table:multi_target}  compares the performance of our models with the baselines for the multiple  target extraction task. It shows  that our method outperforms the baselines by 1.2-1.4 dB for the 2-target case and 1-1.2 dB for the 3-target case. As with prior work~\cite{2020arXiv200605712O}, the SI-SNR improvements are  lower in the 3-target selection task since there is greater similarity between the input mixture and the target signal, compared to the single-target  case, resulting in a larger input SI-SNR. {We obtained $p$-values $< 0.05$ for all comparisons except for the comparison between $(E=512; D=256)$ and $(E=256; D=256)$, for which the $p$-value was $0.57$.}

{
In Fig.~\ref{fig:qual}, we qualitatively show an example of single-target extraction and multi-target extraction from a 4-class input mixture, using our multi-target extraction model. Fig.~\ref{fig:qual_mixture} shows the input mixture waveform, and Figs.~\ref{fig:qual_gt_bark} and \ref{fig:qual_gt_cough} show the isolated ground-truth sounds. We provide the input mixture to our multi-target model, with a single-target query followed by a two-target query. Figs.~\ref{fig:qual_output_single} and \ref{fig:qual_output_multi} are the output waveforms obtained when the single target and the two targets are queried, respectively. The waveforms show that the model successfully recognizes the queried events and extracts the relevant sounds. It can also be observed that our model preserves the original amplitudes of the sounds in the input mixture well.
}

{We also implemented a non-causal version of the proposed Waveformer. The dilated causal convolution (DCC) block was made non-causal by padding on both sides of each DCC layer’s input sequence, while the causal version only padded to the left. The non-causal transformer decoder attends to the previous and next chunks, in addition to the current chunk. Following \cite{2020arXiv200605712O}, we trained our non-causal model with both one-hot/multi-hot based extraction and Permutation Invariant Training + Oracle Selection (PIT + OS) objectives.  Table \ref{table:non_causal} compares the performance of the non-causal version of our model with non-causal baselines. The performance of our causal model is only 1.1 dB less than the non-causal version, in contrast, to the 3--5 dB observed in prior source separation works \cite{luo2019conv, subakan2022resource}. We achieve this resilience by using layer normalization throughout our architecture (avoiding the gLN to cLN switch in \cite{luo2019conv,subakan2022resource}) and a small context length for the transformer decoder.}





\begin{table}
\label{non_causal}
\caption{{Performance comparison with non-causal baselines.}}
\vspace{-.8em}
\centering
\begin{tabular}{lccc}
\toprule
Model                   & SI-SNRi  \\
\midrule
Listen to What You Want \cite{2020arXiv200605712O} & 9.91 \\
Ours ($E = 512;~D = 256$; Non-causal) & 10.50 \\
Ours ($E = 512;~D = 256$; Non-causal; PIT+OS) & 11.31 \\
\bottomrule
\end{tabular}
\label{table:non_causal}
\vskip -0.15in
\end{table}
\section{Conclusions}



We demonstrate the first deep learning  method for real-time and streaming target sound extraction. 
Future work includes the use of more constrained computing platforms,  larger datasets with more classes, and multiple microphones.  Our  Waveformer architecture may be applicable to other   acoustic  applications like source separation and directional hearing, which deserves further exploration.

\vskip 0.05in\noindent{\bf Acknowledgements. } The UW researchers are funded by the Moore Inventor Fellow award \#10617 and the National Science Foundation.


\bibliographystyle{IEEEbib-abbrev}
{\footnotesize
\bibliography{refs}
}
\end{document}